\documentclass[conference]{IEEEtran}
\IEEEoverridecommandlockouts
\usepackage{cite}
\usepackage{amsmath,amssymb,amsfonts}
\usepackage{algorithmic}
\usepackage{algorithm}
\usepackage{graphicx}
\usepackage{textcomp}
\usepackage{enumitem}
\usepackage{xcolor}
\usepackage{multirow}
\usepackage{url}
\def\BibTeX{{\rm B\kern-.05em{\sc i\kern-.025em b}\kern-.08em
    T\kern-.1667em\lower.7ex\hbox{E}\kern-.125emX}}
\begin{document}

\onecolumn
This is a pre-print of the following chapter: Sanil S. Gandhi, Arvind W. Kiwelekar, Laxman D. Netak, Hansraj S. Wankhede; \textquotedblleft Security Requirement Analysis of Blockchain-based E-Voting Systems\textquotedblright, published in Intelligent Communication Technologies and Virtual Mobile Networks, 2022, edited by G. Rajakumar, Ke-Lin Du, Chandrasekar Vuppalapati, Grigorios N. Beligiannis,  2022, Publisher Springer Singapore. The authenticated version is available on publisher site. \\ \\
\textbf{Cite this chapter as:} Gandhi, S.S., Kiwelekar, A.W., Netak, L.D., Wankhede, H.S.(2023). Security Requirement Analysis of Blockchain-Based E-Voting Systems. In:Rajakumar, G., Du, KL., Vuppalapati, C., Beligiannis, G.N. (eds) Intelligent Communication Technologies and Virtual Mobile Networks. Lecture Notes on Data Engineering and Communications Technologies, vol 131. Springer, Singapore. \url{https://doi.org/10.1007/978-981-19-1844-5_6}.

\twocolumn
\title{Security Requirement Analysis of Blockchain-based E-Voting Systems\\
}
\author{
\IEEEauthorblockN{Sanil S. Gandhi, Arvind W. Kiwelekar, Laxman D. Netak, Hansraj S. Wankhede}
\IEEEauthorblockA{\textit{Department of Computer Engineering} \\
\textit{Dr. Babasaheb Ambedkar Technological University}\\
Lonere, Raigad-402103, India \\ ssgandhi@dbatu.ac.in, awk@dbatu.ac.in, ldnetak@dbatu.ac.in}
}

\maketitle

\begin{abstract}
In democratic countries such as India, voting is a fundamental right given to citizens of their countries.  Citizens need to physically present and cast their vote in ballot-paper based voting systems. Most of the citizens fail to fulfill this constraint and have stayed away from their fundamental duty. Electronic-voting systems are often considered as one of the efficient alternatives in such situations. Blockchain Technology is an emerging technology that can provide a real solution as it is characterized by immutable, transparent, anonymous, and decentralized properties. This paper presents a  security requirement analysis for e-voting systems and evaluates blockchain technology against these requirements.
\end{abstract}

\begin{IEEEkeywords}
Blockchain Technology, E-Voting, Cryptography, Security Analysis.
\end{IEEEkeywords}

\section{Introduction}
In Democratic countries, an election plays a vital role in the selection of the government of the respective country. A voter must cast the vote securely and privately without interference from any political party's agents. There are enormous ways to cast the vote in the different countries. In the traditional voting system, a paper is used to cast the vote. The drawbacks of this system are invalid votes, printing of millions of ballot papers, transportation, storage and distribution of ballot papers, stealing or altering a ballot boxes, and the counting process is manual, which takes too long time.

In India, the Postal Ballot system facility is available for the members of the armed forces, members of the state police department, or any person who is appointed on the election duty. Voter have to punch his / her choice on the ballot paper and, then these votes are dispatched to the Returning Officer of the respective assembly using postal services. The drawback of this system is, sometimes ballots are not delivered on time, papers tore in between the transportation or did not properly punch by voters lead to the cancellation of the vote at the time of the vote counting process.

\textit{Electronic Voting Machine} (EVM) \cite{evm} drastically changes this scenario in India. It helps to overcome all the drawbacks of a paper-based voting system. But in this, the voter has to visit the polling station on election day. Another way to cast a vote is through an Internet-based or remote voting system. As voter can cast vote from anywhere, leads to an increase in voter participation. The system is designed as a web-based applications or as a mobile applications.  

Earlier, the web-based applications are connected with bio-metric devices, cards, passwords or PINs, etc. and currently, mobile-based applications have in-built functionality for voter authentication. The online voting system was designed in 2003, by the Estonian government for the local government council elections, where a National ID card with Public Key Infrastructure (PKI) used to authenticate the voter \cite{a1}. P. Y. A. Ryan et at. \cite{pret} have been implemented \emph{Pr\^et \`a Voter}, a secure electronic voting system with end-to-end verifiability.

Such, Internet-based systems are centralized, and vulnerable to many attacks as well as depended on a trusted third party to carry out the whole election process. Internet-based systems are easy to penetrate and are less transferable as there is possibility of data manipulation. Due to such issues in the existing online voting systems, many researchers are working on the development of full-proof remote voting systems using innovative and emerging technologies such as blockchain technology. The primary design goal of this kind of efforts is to build a voting system which shall preserve voter’s privacy and it shall be resilient to anticipated and unforeseen security threats.

Some of the platforms for mobile-based remote voting, such as Voatz, Votem, etc. are recently used to conduct elections for citizens having physical disabilities, busy in their daily routine, or leaving out of the city. \textit{Voatz} is used within certain jurisdictions where local US government offers their citizens to remotely cast the vote in the election. Private election of the organisation is also conducted. For example, the Ohio State Bar Association election conducted using the \textit{Votem} blockchain-based mobile voting platform \cite{votem}.

Follow my votes, TIVI, BitCongress, and Vote Watcher are the examples of the blockchain based mobile and desktop applications. The details of the such applications as follows:
\begin{enumerate}
    \item \textbf{Follow My Vote \cite{fol}:} It is the online, open-source voting platform with end-to-end verifibility. Desktop application is working on live operating system, and mobile based application needs the various administrative permissions at the time of installation of application so there is less possibility of malware attacks.
    \item \textbf{TIVI \cite{tivi}:} TIVI is remote voting application with easily understandable user interface to novice users available for all the devices. This application ensures the voter identity and eligibility through authentication methods. Every ballot is secured using the public key encryption as well as voter's digital signature. Zero-Knowledge Proof \cite{zkp} is used for correct voting process.
    \item \textbf{BitCongress \cite{bit}:} BitCongress is the peer-to-peer electronic voting platform. The platform is designed using Bitcoin, Smart Contract, Counterparty, and Blockchain APIs. Proof-of-Work, Proof-of-Tally such mechanisms are used for data integrity.
    \item \textbf{VoteWatcher \cite{vote}:} It is the open source electoral system. The `Optical Mark Recognition' (OMR) sheet is used as a paper ballot. Data extracted as voter cast the vote and stored on the blockchain.
\end{enumerate}

In the paper \cite{a2}, authors analyzed two e-voting systems. First one is the \textit{Estonian E-Voting System} (EstEVS) and another is the \textit{Secure Electronic Registration and Voting Experiment} (SERVE) to find out which is secured. The authors proposed the game-theoretical methods to analysed the large-scale attacks. It is based on the probabilities model. The required data for analysis of this system is collected from 1000 machines. The possible large-scale attacks in such systems can be large scale votes’ theft, disenfranchisement of votes, votes buying and selling, and privacy violation.

In the next section II, we discuss the blockchain primer and related work in the blockchain based e-voting systems. Section II presents the detailed requirement analysis of the blockchain-based e-voting system. Finally, we conclude this paper in section IV.

\section{Background and Related Work}
\subsection{Blockchain}

\begin{figure}[hbpt]
\centering
\centerline{\includegraphics[width=0.5\textwidth]{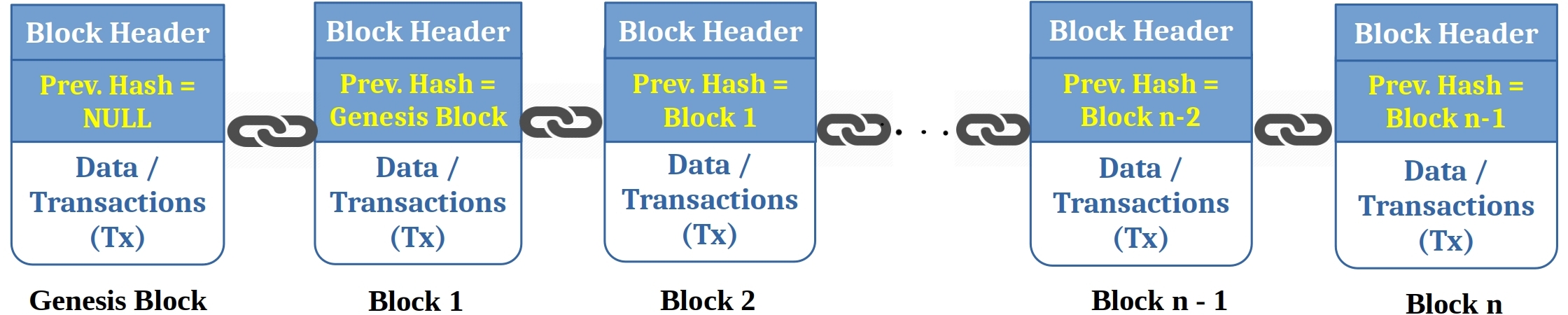}}
\caption{Blockchain}
\label{fig1}
\end{figure}

Bitcoin (Cryptocurrency) \cite{btc} is the first application designed using the blockchain platform. Blockchain stores data in the forms of the chain of blocks as shown in Figure \ref{fig1} in append-only mode. If, adversary tries to alter or delete the data/transactions stored in the block, it violates the consensus rules. Number of transactions are stored in a block as per the maximum size of the block.
The first block of blockchain is called a \emph{genesis block}. In genesis block, the previous hash field is set to a default value in the genesis block means it sets to zero or null and for the remaining fields, values are set to default. The key characteristics of blockchain technology such as decentralised, immutable, anonymity, and transparency are described as follows: 
\begin{enumerate}
    \item \textbf{Decentralized:} The single-point failure, collapse the whole centralized system leads to the decentralization of the data. In this, data is not controlled by a single entity. As, it forms trustworthy platforms, no need of a third party or inter-mediator when performing transactions with unknown parties.
    \item \textbf{Immutable:} This property guarantees that no one can able to change the data stored on the blockchain. All the nodes that are part of network stored the data. An eavesdropper alters or deletes the data on the blockchain if and only if an eavesdropper has more than 51\% control in the system.
    \item \textbf{Anonymity:} This property hides the original identity of the user from the rest of the world. It is useful in fund transfer or e-voting activities where the needs of hiding the identity of the person. This is achieved via public/private key-pair addresses.
    \item \textbf{Transparency:} All the blocks consisting of data is stored in every node present in the network. So, any node that joins the e-voting system can able to verify or view every transaction. 
\end{enumerate}
These key elements help to design a full-proof and efficient e-voting system. It removes most of the pitfalls that arose in the traditional and Internet-based i.e. centralized e-voting system. 

Blockchain implementation is based on the Distributed Ledger, Cryptography, Consensus Protocol, and Smart Contract.
\begin{enumerate}
    \item \textbf{Distributed Ledger (DL):} Traditional ledgers are used in business for keeping day-to-day accounting information, the same way in blockchain distributed ledgers store all the transactions in all the peers that are part of the network. Any alteration in the ledger leads to an update in all the consecutive nodes. Voter's information, votes cast by voters are recorded and replicated on every ledger.
    
    \item \textbf{Cryptography (Crypt.):} Paillier encryption \cite{bb9}, Homomorphic encryption \cite{bb10}, Mix-net \cite{b10}, Ring signature \cite{b3}, Blind Signature \cite{b11}, as well as Zero-Knowledge Proof (ZKP) \cite{zkp} etc. cryptographic and encryption techniques prefer to secure any systems. These techniques are used for user authentication, message privacy by encrypting and decrypting the data and hiding the user's identity by a key-pair generated using asymmetric cryptographic algorithms.
    
    \item \textbf{Consensus Protocol (CP):} All nodes in the decentralized network reach a common agreement regarding the current state of the ledger with the help of a consensus algorithm. It eliminates the double-spending attack. Bitcoin \cite{btc} uses Proof-of-Work (PoW), Ethereum shifted from PoW to Proof-of-Stake (PoS), and Hyperledger prefers Practical Byzantine Fault Tolerance (PBFT) as a consensus algorithms. As per business specifications, various consensus algorithms such as Proof of Activity (PoA), Proof of Weight (PoW), Proof of Importance (PoI), Delegated Proof of Stake (DPoS), Proof of Elapsed Time (PoET), Proof of Capacity (PoC), etc. can be used.
    
    \item \textbf{Smart Contract (SC) \cite{smt}:} Like the legal agreement or contract signed between two parties, to exchange some value as parties satisfy the constraint mentioned in the agreement. In the same manner, a smart contract triggers a particular event when the system meets the predefined condition specified in the contract. Solidity, Vyper, Golang, etc. are some of the programming languages to write smart contracts. In e-voting, smart-contracts can be used to convey a message to a voter after the vote stored on the blockchain, carry out election rules, keep the track of the start and end of election time, etc.
\end{enumerate}
There are different e-voting systems are designed using the blockchain platform. That can be classified as first is based on public or private cryptocurrencies such as Bitcoin \cite{btc}, Zcash \cite{b12}, Zerocoin \cite{a4}, etc.; the second uses the smart contract to trigger events as met with defined condition, and last is by making use of blockchain as ballot boxes \cite{b3}.

\subsection{Related Work}
Zhao Z. et al. \cite{a3} proposed the bitcoin-based e-voting system. In this system, candidates have to deposit bitcoins on every voter's account before the start of the election. This system does not support large scale voting. In the paper\cite{b1}, the authors used the Homomorphic ElGamal encryption and Ring Signature methods. The advantage of the system is that transactions performed by the voters are confirmed in less time. Authors\cite{b2} proposed the blockchain-based solution. It uses the cryptographic blind signatures technique. The pitfalls of this system is that ballots are not encrypted after the vote cast by the voter and it is possible to trace the IP address of voters and nodes using the network analysis tools.

B. Yu et al.\cite{b3} suggested various cryptographic techniques such as Paillier encryption, Ring signature, Zero-Knowledge Proof (ZKP) to efficiently run the voting process. Anonymity is achieved through a key-pair of 1024 or 2048 bits generated for every voter using short linkable ring signature technique. The only downside of ZKP, if a voter forgets his/her passcode, he/she lose the data forever. In the paper \cite{b4}, the authors explained how blockchain technology makes e-voting secure, but the authors have not implemented the e-voting system using the blockchain platform.

A private blockchain-based e-voting system proposed by F. Hj\`{a}lmarsson et al.\cite{b5}. Authors \cite{b6} suggested the Fake Password technique, which is the combination of fake credentials \cite{b7} and panic password \cite{b8} to avoid voter coercion. To avoid Danial-of-Service (DoS) attacks, authors used Challenge-Response Protocol, e.g. CAPTCHA, at the time of the creation of fake passwords by an adversary and also data servers to be distributed across the network to store the ballots during the voting process. In this paper, the author did not mention how many votes a voter can cast during the voting process.

In the paper \cite{board}, the author proposed a blockchain-based e-voting system implemented using ethereum. It is the internet-based decentralized system which is capable of self-tallying the votes in the counting phase, provides voter privacy, and end-to-end verifiability. But, the system is not able to scale for more than 50 voters and also system is vulnerable to Denial-of-Service attacks. Dagher G. et al. \cite{bron} proposed the ethereum blockchain-based system for conducting the students' association elections at the university level. The authors suggested homomorphic encryption techniques, cryptographic algorithm to achieve the voter's privacy.

Hardwick et al. \cite{b9} proposed a private, blockchain-based e-voting system implemented using the \textit{Ethereum} platform. The voter can alter his/her already casted vote number of times till the end of vote casting phase. But, when the counting process starts, only the voter's last vote is considered for counting. The downsides of the system are (i) There is need of the Central Authority (CA) to verify the voter authorization; (ii) A voter can alter the vote till the election end time. To store multiple votes from a voter, there is a need of an extra storage space, and the cost of computation to eliminate duplicate votes is too high; (iii) To count the vote in the counting phase, voter receive the ballot opening message. This message is required to open or count the vote stored on the system. Because of any issue, if voter unable to send a ballot opening message to the counting authority, then his/her vote will not be consider in the final tally.

\section{Requirement Analysis}
Every eligible voter must participate in the election process, and Election Commission (EC) tries to make available various reliable, trustworthy, and secure platforms so no one can stay away from their fundamental rights. The new voting platforms must have to satisfy all the legal constraints of the existing and traditional voting process. The complete election process is separated into several tasks. Dimitris \cite{b14} elaborated the system requirements for secure e-voting. Software Engineering principles assist in identifying these functional and non-functional requirements of the e-voting process.

\subsection{Functional Requirements (FR)} 
In this, we are specifying the various services or functionalities that are offered by the system. These functional requirements specify the inputs and their behavior as per the inputs provided by the user to the system. The abstract view of all e-voting related functional requirements are described as given below:
\begin{enumerate}
    \item \textbf{Voter Registration [FR1]:} Every voter has to register herself in front of election authority with necessary documents. The screening criteria help to identify the eligibility of the voter.
    \item \textbf{Provide the Authentication Credentials to Voter [FR2]:} Each voter, after validation of the identity of voter, gets credential details in the form of either user-id and password or public-private key pair. The credential hides voters' real identity from the rest of the world. At the time of vote casting, every voter authenticated using credentials and this will help to prevent double-voting.
    \item \textbf{Prepare the Digital Ballot Paper [FR3]:} As the candidates are finalized for the election, the next phase is to prepare the ballot paper to cast the vote. The ballot paper consists of the name of all the candidates. As the start of vote casting phase, every voter received a ballot paper to cast a vote via email or notification on client-side software.
    \item \textbf{Casting the Vote [FR4]:} Each voter, select one of the candidates from the list of candidate mentioned on the ballot paper. No link between voter and vote is available to prove the way a vote is cast by a voter. For this, each vote must be encapsulated by a double layer. First by using the voter's digital signature and then vote encrypted by one of the encryption techniques. The votes cast by voters are verified against the already available votes to avoid the double-voting problem. After verification, a vote is stored on the blockchain. A voter can cast a vote in a stipulated duration of time. 
    \item \textbf{Vote Tallying [FR5]:} After the end of the vote casting phase, the system starts to count the votes stored on the blockchain, after counting the result is stored on the blockchain for the audit purpose, and then declare the results of an election. In this phase, all the votes are fetched from the blockchain, then read the field by decrypting every vote and added it into an individual account.
\end{enumerate}
As per the various constraints mentioned by Election Commission for each functional requirements, we mapped these functional requirements to the blockchain architectural elements are shown in the Table \ref{map_FR} given below:

\begin{table}[htpb]
\caption{Mapping of System's FR to Blockchain Elements}
\label{map_FR}
\begin{tabular}{l|c|c|c|c|l}
\hline
\multirow{2}{*}{\begin{tabular}[c]{@{}l@{}}Functional\\ Requirements\end{tabular}} & \multicolumn{4}{l|}{Blockchain Elements} & \multicolumn{1}{c}{\multirow{2}{*}{Remark}} \\ \cline{2-5}
 & \multicolumn{1}{l|}{DL} & \multicolumn{1}{l|}{Crypt.} & \multicolumn{1}{l|}{CP} & \multicolumn{1}{l|}{SC} & \multicolumn{1}{c}{} \\ \hline
\begin{tabular}[c]{@{}l@{}}Voter\\ Registration \\ {[}FR1{]}\end{tabular} & x & x &  & x & \begin{tabular}[c]{@{}l@{}}It records a hash of\\ data in the DL. The\\ consensus protocol\\ shall ensure consist-\\ ency among all\\ copies of DL.\end{tabular} \\ \hline
\begin{tabular}[c]{@{}l@{}}Provide the\\ Authentication\\ Credentials to\\ Voter {[}FR2{]}\end{tabular} & x & x &  &  & \begin{tabular}[c]{@{}l@{}}Assign unique key\\ using public-private\\ key generation \cite{yak}.\end{tabular} \\ \hline
\begin{tabular}[c]{@{}l@{}}Prepare the\\ Digital Ballot\\ Paper {[}FR3{]}\end{tabular} & x & x &  & x & \begin{tabular}[c]{@{}l@{}}Implement the logic\\ of automated generation\\ of ballot paper as a smart\\ contract.\end{tabular} \\ \hline
\begin{tabular}[c]{@{}l@{}}Casting the\\ Vote {[}FR4{]}\end{tabular} & x & x & x & x & \begin{tabular}[c]{@{}l@{}}Distributed ledgers\\ used to record the\\ vote. The consensus\\ protocol shall ensure\\ consistency among\\ the multiple copies\\ of ledgers.\end{tabular} \\ \hline
\begin{tabular}[c]{@{}l@{}}Vote Tallying\\ {[}FR5{]}\end{tabular} & x & x &  & x & \begin{tabular}[c]{@{}l@{}}Implemented as a\\ smart contract to count\\ the vote received by\\ each candidate.\end{tabular} \\ \hline
\end{tabular}
\end{table}

\subsection{Non-Functional Requirements}
Non-functional requirements define systems quality attributes that affect the user experience. If unable to fulfill the non-functional requirements, then the system will not satisfy the user's expectations. The most of the systems are designed for the small scale election but not for the large scale election. Zhang et al. \cite{chain} implemented the blockchain-based e-voting system for large-scale elections. In this, author evaluated the performance of the Chaintegrity based on nine different non-functional requirements with the help of numerous cryptographic algorithms and encryption techniques.

There are several non-functional requirements but, in this we are targeting security related quality attributes. E-voting is a confidential process and must protect against security attacks, such as Sybil attack, Denial-of-Service (DoS) attack, Man-in-the-middle attack, etc. These non-functional requirements are listed as a given below:

\begin{enumerate}
    \item \textbf{Privacy:} This requirement satisfies that no one (including the electoral authorities) can be traced out to whom the vote is cast by the elector. It is not possible to reveal the candidate chosen by the voter in the vote casting phase, and all votes remain concealed. Voter name, elector’s public/private key-pair, voters' documents, etc., will not be shared with unauthorised users.
    \subitem\emph{Anonymity} means voters can cast a vote with a pseudonymous identity using the public key. This can be achieved through cryptographic algorithms such as Blind or Ring Signature.
    \subitem In \emph{receipt-freeness} \cite{recpt}, a receipt will not be generated after a vote cast by an elector. It ensures that voters will not be able to prove to coercer to whom they voted. Even if the opponent cracked the secret key of the elector, he would not find out the elector's vote choice.
    \subitem\emph{Coercion-resistance} means the system must be designed in such a way that an adversary can not force a voter to cast a vote to a specific candidate, or voters can not prove to an adversary how they voted.
    
    Both receipt-freeness and coercion-resistance are complementary to each other.
    
    \item \textbf{Data Integrity:} The adversary should not access or tamper with the data stored on the system. Also, unauthorized users should not alter the vote cast by electorates and the electorates’s personal information. Cryptographic hash algorithms are one-way functions, so the integrity of data can remain intact.
    
    \item \textbf{Confidentiality:} means the system will not reveal the true identity of the electorate to their public-key infrastructure. An authorized user may have access to the system data. This confirms that the system has quality attributes such as coercion-resistance and privacy.
    
    \item \textbf{Authentication:} The legal and eligible voters only access the system and cast the vote, and it helps in preventing the system from data or information leakage.
    
    \item \textbf{Authorization:} Different stakeholders of the system are provided with a wide range of functionalities. These functionalities help in preventing illegal access and tampering with the data.
    
    \item \textbf{Auditability:} After the counting process, any external auditors can recount the votes stored on the blockchain. Also, voters verify whether their vote counted in the final tally or not. This property helps to achieve the verifiability and validity of the process.
    
\end{enumerate}
There are also other non-functional requirements such as verifiability, scalability, transparency, availability, reliability, etc., need to improve the e-voting system's performance. Usability and accessibility are non-functional system requirements that also play a crucial role in making the system successful. To satisfy these constraints, developers must select the public, private, or consortium blockchain networks as per the need in the system requirement.

\begin{table}[htbp]
\caption{Non-functional Requirements of Existing Blockchain-based E-Voting Systems}
\label{tab:NFR}
\begin{tabular}{l|c|c|c|c}
\hline
\multicolumn{1}{c|}{\multirow{2}{*}{\begin{tabular}[c]{@{}c@{}}Non-functional\\ Requirements\end{tabular}}} & \multicolumn{4}{c}{Existing Blockchain-based Platforms} \\ \cline{2-5} 
\multicolumn{1}{c|}{} & FollowMyVote & TIVI & BitCongress & VoteWatcher \\ \hline
Privacy               & x            & x    & x           &  \\ \hline
Integrity             & x            & x    & x           & x \\ \hline
Confidentiality       & x            &      & x           &  \\ \hline
Authentication        & x            & x    &             &  \\ \hline
Auditability          & x            & x    & x           & x \\ \hline
\end{tabular}
\end{table}
 
Table \ref{tab:NFR} maps the non-functional requirements of e-voting system to the some of the existing commercial Blockchain-based voting applications such as FollowMyVote \cite{fol}, BitCongress \cite{bit}, and TIVI \cite{tivi}, VoteWatcher \cite{vote}.

\subsection{Countermeasures on the Typical Attacks}
Adversary continuously trying to damage the existing system using different potential attacks. Blockchain technology combined with some other techniques such as authentication makes the system robust against the vulnerabilities and helps to minimize attacks. In this section, we are elaborating on the various attacks in the e-voting system and countermeasures.
\begin{enumerate}
    \item \textbf{Denial of Service (DoS) Attack:} In DoS, attackers target weak nodes in the network by continuously sending malicious packets to flood the network with excessive traffic. These attacks are resisted by using various techniques such as authentication, Challenge-Response protocol \cite{chall} e.g. CAPTCHA, etc. In a blockchain, a distributed ledger prevents DoS attacks since data or services are distributed on multiple nodes.
    \item \textbf{Man-in-the-Middle Attack:} In a Man-in-the-middle attack, an adversary intercepts communication between two parties to tamper the messages. An adversary uses the same transaction and re-transmits the same data to the smart contract. Various methods such as validating every single vote presets on the block before the block is stored on the blockchain along with the voter's public key after casting a vote are preventing such kind of attack.
    \item \textbf{Replay Attack:} This attack is the same as the Man-in-the-Middle attack. An adversary tries to copy the original transactions from one blockchain along with the user's signature and performs a new transaction on the blockchain. In the case of e-voting, it is the same as double voting. Using the random key/token for a every single transaction, double voting avoiding strategies, or current timestamp of the transaction, etc. can be used to prevent such attacks from a malicious user. 
    \item \textbf{Sybil Attack:} In this, a malicious user will create multiple accounts or runs multiple nodes in the distributed network. Using permissioned blockchain platforms or voter authentication functionality to allow only eligible voters to cast the vote can defend against such attacks.
\end{enumerate}
Basic building blocks of blockchain architecture helps in prevent and mitigate the numerous attacks such as smart contract-based, consensus protocol-based, mining-based, or wallet-based attacks. 

\subsection{Security and Privacy Requirements of Voting}
The following are some of the generic security and privacy requirements that any voting system shall enforce.
\begin{enumerate}
    \item The voting system shall able to preserve the privacy of the voting information, means it shall not disclose the information that who has casted a vote to which candidate.
    \item The voting system shall able to preserve the privacy of ownership of ballot means it shall not disclose the identity of voter associated to a particular ballot. This is also known as anonymity of voters.
    \item The voting system shall prove to the voters and auditors that no ballots are removed from the ballot box.
    \item The voting system shall be coercion resistant means that voting system shall protect voters from casting a vote under threat or attack.
    \item The voting system shall not be able to tamper with the casted votes once election is closed and in-process.
    \item No early results should be obtainable before the end of the voting process; this provides the assurance that the remaining voters will not be influenced in their vote.
\end{enumerate}


\section{Conclusion}
The design of blockchain-based voting system for a large-scale of election must be secure, reliable, trustworthy, robust and free from security loopholes. The system's workflow must be user-friendly, and the system is easily access or operate by the masses. Still, remote voting using blockchain for huge population is under development phase and needs to find out various techniques for coercion-resistance, and smoothly conducting the voting process.

This paper presents functional and non-functional requirements for blockchain-based remote e-voting systems. Also, these functional and non-functional requirements are mapped to the basic architectural elements of the blockchain. Some of existing blockchain-based voting systems are also reviewed. Hence, paper presents state of the research in the blockchain-based e-voting systems, and how to implement countermeasures to overcome the various vulnerabilities.


\vspace{12pt}

\end{document}